# End-to-end verifiability

Josh Benaloh, Microsoft Research. benaloh@microsoft.com

Ronald Rivest, MIT. rivest@mit.edu

Peter Y. A. Ryan, University of Luxembourg. peter.ryan@uni.lu

Philip Stark, UC Berkeley. stark@stat.berkeley.edu

Vanessa Teague, University of Melbourne. vjteague@unimelb.edu.au

Poorvi Vora, George Washington University. poorvi@gwu.edu

February 2nd, 2014.

This pamphlet describes end-to-end election verifiability (E2E-V) for a nontechnical audience: election officials, public policymakers, and anyone else interested in secure, transparent, evidence-based electronic elections.

This work is part of the Overseas Vote Foundation's End-to-End Verifiable Internet Voting: Specification and Feasibility Assessment Study (E2E VIV Project), funded by the Democracy Fund.[1]

## Introduction – what is end-to-end verifiability?

Getting the election outcome right isn't good enough. Voters deserve convincing evidence that the outcome is correct.

There are many ways an election can go wrong or be suspect. Perhaps ballots were lost—and known to be lost. Perhaps a memory card failed—and was known to have failed. Perhaps more votes were reported than there were voters. Perhaps the announced outcome is correct but the evidence supporting it is incomplete or unconvincing. Perhaps some people claim to have observed irregularities. *An election might appear outwardly to have been conducted properly but have an inaccurate outcome due to undetected errors or fraud.*

Well-managed paper-based elections include administrative mitigations to avoid these failures using a combination of human processes and physical evidence. If polling and counting processes are transparent, observers can watch the polling place throughout the election and scrutinize the count afterwards. Rigorous auditing conducted transparently and under observation can reinforce or correct the announced outcome.

Even under ideal circumstances—better than can be expected in real elections—typical voters have no eyes on the process. From their perspective, they cast their votes, go home, and are told the outcome. If they distrust election officials, equipment, or processes, there is little that they can do.

*End-to-end verifiable election techniques enable individual voters to check crucial ingredients of election results – without requiring voters to trust election software, hardware, election officials, procedures, or even observers. Voters may check these ingredients themselves, place their trust in others of their choice (e.g. their preferred candidates, news media, and/or interest groups), or accept the outcome produced with the usual administrative safeguards.*

---

[1] Thanks to Judy Murray and Keith Instone for helpful comments on an earlier draft.

## Comparison with other software independent voting methods

E2E-V is not the only sensible option for computerized voting. In many cases some other software-independent system might be more appropriate. Software-independent systems, as defined by Rivest and Wack (Rivest 2008), are voting systems where an error in the voting system software will not cause an undetectable error in the election outcome. In such a system, one need not trust the voting system software to be correct in order to accept the election outcome, because there are other chains of evidence that can help check that the software performed its task correctly.

Most jurisdictions in the USA have accepted that electronic records from paperless DREs are not good evidence, motivating a return to permanent paper records. New techniques for risk-limiting audits have convinced most people that a properly conducted random audit is a good (and much cheaper) substitute for a full manual count in many circumstances. But audits still rely on the integrity and completeness of the paper records that are audited—which itself requires evidence. The integrity of the audit trail may be difficult to guarantee. It may be verifiable (for ballots cast in person) or not (for remote voting). Verifiably secure transport and storage of paper records is particularly challenging for remote voting, whether in a supervised or unsupervised polling place.

Any chain of evidence, for any system design, requires some assumptions. Checking that those assumptions hold for the election in question is crucial. Trusting unverifiable hardware or software is generally not reasonable. Trusting processes for secure handling of paper ballots may, or may not, be reasonable, depending on what those processes are and how they can be observed. E2E-V systems are software independent and do not rely on trusted paper processes, instead providing opportunities to verify the election outcome electronically.

Other considerations such as usability, accessibility, resilience, vote privacy and resistance to coercion are also important, but not part of this document.

## How to achieve end-to-end verifiability (E2E-V)

End-to-end verifiability is a collection of techniques for replicating, and in some ways exceeding, the standards of evidence provided by an ideal, observed, polling place. This includes two principal components.

1. *Cast As Intended:* voters can verify that their selections (whether indicated electronically, on paper, or by other means) are correctly recorded, and
2. *Tallied As Cast:* any member of the public can verify that every recorded vote is correctly included in the tally.

Consider first a simple way to conduct a verifiable election: the names of voters along with their votes are posted on a public "bulletin board" of some kind. Voters can check the bulletin board to see whether their own votes appear correctly, and everyone can confirm the tally of the listed votes. This election tally is easily verifiable by anyone[2] and requires no trust in election officials, but there is no privacy – all votes are public.

So, let's try a more sophisticated approach: an election authority assigns each voter a pseudonym. After each voter votes, the authority posts the pseudonym next to the vote. Again, voters can individually check that their correct votes appear alongside their respective pseudonyms. This protects ballot privacy somewhat (some voters might accidentally or deliberately reveal their pseudonyms) but it still has problems – for example, voters are susceptible to coercion, and the authority might cheat, for instance by assigning the same pseudonym to voters who are likely to vote the same way.

---

[2] Voters must trust that *other* voters are also checking that *their* votes appear correctly.

So, we move to a still more sophisticated approach. Each vote is encrypted before it is cast, meaning that it can be read only by someone (such as the electoral authority) who holds the secret decryption key[3]. The authority posts each encrypted vote on the bulletin board next to its voter's true name. Now of course, to compute the outcome, the authority has to compute the tally from encrypted votes in a way that will preserve ballot privacy, and to check the outcome, voters need to be able to verify that the tally of encrypted votes is correct. In most end-to-end verifiable systems, the two principle requirements described above (*cast as intended* and *tallied as cast*) are achieved in three phases, described here.

1. *Cast As Intended:* voters make their selections and, at the time of vote casting, can get convincing evidence that their encrypted votes accurately reflect their choices;
2. *Recorded As Cast:* voters or their designees can check that their encrypted votes have been correctly included, by finding exactly the encrypted value they cast on a public list of encrypted cast votes; and
3. *Tallied As Recorded:* any member of the public can check that all the published encrypted votes are correctly included in the tally, without knowing how any individual voted.

End-to-end verifiable voting schemes provide ways of performing these tasks. Consider the *cast as intended* step. One solution is to tell voters that they may use the voting device to produce as many encrypted ballots as they like, but choose only one to cast. All the other encrypted ballots serve as "challenges" and are subsequently decrypted to provide evidence that they match what the voters expect. This is often called a "Benaloh" challenge. Checking that the decryption is honest may require assistance, but these need not be burdensome for voters. While voters could expend the effort to use any of a variety of means to check the decryptions in real time, it is simpler for the election authority to post all challenged ballots on a public list (distinct from the list of cast ballots).

It is crucial that the voting device does not know which encryptions will be challenged. If enough voters select randomly and independently which encryptions to challenge, the machine is very likely to be "caught" if any significant portion of its encryptions are faulty. Also importantly, it is not necessary for all voters to challenge encryptions for this detection probability to be large. The crucial assumption is that the voting device cannot predict with confidence that certain voters will not challenge.

The Helios E2E-V Internet voting system ([vote.heliosvoting.org](vote.heliosvoting.org)) was one of the earliest systems to use this approach. The Wombat voting system, [http://www.wombat-voting.com/](http://www.wombat-voting.com/), and StarVote both combine this sort of challenge mechanism with a paper evidence trail.

Prêt à Voter and Scantegrity II use a related approach, but the encrypted candidate names are generated before voting and printed on paper alongside human-readable candidate names. Each voter marks a selection of those (encrypted) candidate names. Interested voters can challenge printed ballots before they vote, which means checking that they are "correctly constructed" and hence would result in correct encryptions of the vote.

The *recorded as cast* step usually involves giving data about the encrypted ballot to the voter as a "receipt" when the voter casts the ballot. Voters may subsequently check the published list to ensure that the data on their receipts is published correctly.

The *tallied as recorded* step uses cryptographic mathematical proofs, usually in one of two ways. The first method, called "verifiable mixing," disassociates the votes from voters' names by "shuffling," then decrypts the votes.  This shuffling is performed in a way that allows the authorities to prove that none of the votes have been altered but does not allow the public to trace a given vote to any particular voter.  The second method, which uses "homomorphic encryption," makes it possible to tally the encrypted votes without decrypting them and then prove that the tally is correct.  Both methods produce a mathematical proof that the announced outcome matches the published encrypted cast votes.

---

[3] The encryption step also adds some randomness, so that two encryptions of the same vote don't look the same.

## What procedures are necessary to support E2E-V?

End-to-end verifiable systems require unfamiliar (and sometimes complex) procedures, which pollworkers must be trained to follow exactly.  If the procedures are not followed to the letter, it may be impossible to give convincing evidence that the announced election result is correct.

## What trust assumptions remain?

End-to-end verifiability puts powerful auditing capabilities into the hands of voters, but it adds nothing if these capabilities are not exercised.  A verifiable election that nobody bothers to verify does not produce any meaningful evidence, and the evidence can be unconvincing if an insufficient number of voters avail themselves of their new capabilities.  Fortunately, in most cases only a tiny fraction of voters need to perform these additional tasks to produce compelling evidence of the correctness of the election outcome.  The additional procedures that are available vary among end-to-end verifiable systems, but they can place an extra burden on voters and election officials.

E2E-verification hinges on the unpredictability of voters.  If the system makes many errors—either encrypting votes incorrectly or losing encrypted votes before they are published—then it does not take many attempted verifications for there to be a large chance of discovering at least one error, provided voters decide whether to verify the encryption of their votes or the presence of their votes on the bulletin board as if at random, independently.  Some verification (such as the voter's opportunity to check that their votes are cast as intended) must be performed on the spot, by the voter, before the election outcome is known.  Other kinds, such as correct recording of votes or a universal check of correct tallying, can be performed afterwards, by anyone.

 (In contrast, methods such as Risk Limiting Audits use public random auditing to develop statistical evidence about election outcomes.  However, this evidence may be convincing to only to voters who observe the audit, or who trust those who observed the audit.)

In current E2E-V systems, much of the verification involves calculations that are beyond the capabilities of humans to perform the checks unassisted.  They need software and computers to perform the checks, but this of course raises a further question – how can we trust the checking software?  The answer is that anyone can, in principle, write software to perform the checks.  Independent organizations (such as interest groups or news media), interested individuals, and even candidates or political parties can provide their own checking software.  Voters and observers can choose any source they trust, use several, or even write their own; in principle, that makes it possible to determine whether the checking software is correct.  More importantly, as long as *enough* checks are performed with sound code, there will be a large chance of detecting errors.  In contrast to traditional systems, where the software is built by vendors behind closed doors and chosen by election officials, with E2E-V systems, the verification process is public and each voter can choose which verification software to trust.

The risk of conflicting results amongst software verifiers is mitigated by public expectations of their results.  A verifier that merely asserts that an election tally is incorrect – without offering specific evidence – can be ignored.  A reasonable verifier should either assert that a tally is correct or offer evidence of an error that is so specific as to be verifiable by hand.  All properly-constructed verifiers should produce the same result.  However, if there is disagreement, even a lone verifier that finds an error in the tally is sufficient to overturn a result if it provides specific and independently-confirmable evidence of an error.

It should be noted that within this context, errors are not trifles.  Instead what is being verified is that a posted set of ballots corresponds to a claimed tally.  Errors at this stage are not caused by misplaced ballots or ambiguous interpretations.  An error here is only possible if the tallying software misbehaves, and in such cases the tally produced could be off by a million votes as easily as off by a single vote.  Confirmed errors need to be rectified to obtain an accurate tally.

## How can you tell whether you've got E2E-V?

Since end-to-end verifiability is recognized as a desirable attribute, some electronic voting software vendors claim they (already) sell it.  Claims are easy, but implementing E2E-V methods is not.  Having examined many commercial products, we know of none that offers true E2E-V.  The crucial properties are the kinds of verification mentioned above, with complete freedom about what software to trust for verification.  ***If voters and election observers do not have complete freedom about which software to trust to do the verification, or if only a privileged few are able to verify at all, the system is not truly E2E-V.***

## Dispute resolution (accountability, non repudiation, defense against defaming).

Our discussion of verification has so far included only a description of what can be verified.  End-to-end verifiability gives voters (collectively) the opportunity to detect errors that might have altered the announced outcome.  We have not discussed procedures for dealing with claims that the system erred.  Of course, claims may be legitimate, but they might also be false.  Hence, we have several closely related questions:

- If voters (or observers) detects that the system erred, how can they prove it?
- What procedures are in place to ensure that credible claims of error are acted on appropriately?
- How can the system (or the authorities) defend against false claims that the system misbehaved?

These questions apply to conventional voting systems too.  For example, a voter could claim that a voter-verifiable paper trail has incorrectly recorded a vote, or an observer of a traditional paper count could claim to have observed misbehaviour.  Because E2E-V presents new kinds of verification, it also introduces new ways to perform this old attack.

A good E2E-V system should specify what happens if a verification step fails.  Authorities and observers should have ways to assess whether a problem is genuine and how much of an election could be affected.  This is called "dispute resolution," meaning that when someone complains about a system malfunction, the accuser can bring evidence to a third party who can resolve the dispute by either accepting the evidence of system misbehaviour or exonerating the system.

For example, many end-to-end verifiable systems provide voters with a paper receipt at the time of voting.  The voter uses this information later to check whether the associated vote was correctly included in the count.  If such receipts are easy to counterfeit, then voters can falsely accuse the system of altering or omitting votes (this is sometimes called a "defaming attack").  Although such attacks cannot manipulate votes undetectably, they can cast doubt on the outcome by creating the impression that someone has manipulated the votes.  Prêt à Voter, and variants such as the vVote system, print a digital signature on each receipt.  This can make it very difficult to forge a paper receipt.  However, a malicious voting system could produce invalid digital signatures on receipts associated with votes that it wishes to discard.  Thus, digital signatures are only effective if voters have independent means to validate the validity of the signature within poll sites (perhaps using a trusted application on a personal smart phone to read a QR code or 2D bar code).  Even then, a protest is only credible if independent entities can observe that the invalid receipt came directly from the election equipment and not, say, the voter's pocket.

As another example, Scantegrity II provides voters with short return codes (usually two letters) which voters only see after committing to their selections. Voters can later use these codes to prove that their votes should have been included. A voter who attempted to defame a system by falsely accusing it of altering or omitting a vote would need to guess the right return code. This is difficult but not impossible – if there were thousands of false accusations, we would expect a handful of correct guesses "by chance." Because the Scantegrity II paper ballots are printed in advance, incorrect return codes would likely be detected during the ballot verification process.

The STAR-Vote system also provides real-time evidence that can be used for dispute resolution. Voters can spoil ballots produced by the system. Upon spoiling a ballot, a voter is in direct possession of both an encrypted receipt and an associated cleartext ballot summary. These can be shown to observers and the paper can be retained by the voter as forensic evidence.

Without dispute resolution, it may not be possible to distinguish genuine errors or manipulation attempts from efforts to defame the system. Hence it may not be possible to verify whether the outcome was indeed correct. However, even without dispute resolution, voters using an E2E-V system will know whether their votes have been properly counted and will know whether the count of the recorded votes is correct. The uncertainty comes if other voters claim that their votes were not correctly included, and external methods must be employed to assess the veracity of any such claims. This uncertainty can be mitigated if a process of real-time auditing is used wherein systems are randomly tested in public with dummy votes; in contrast with normal voters, dummy voters can be observed and even recorded during the process, and any inconsistences can thereby be publically captured.

## Practical deployments of E2E-V

There have been numerous small trials of end-to-end verifiable voting systems. Some interesting polling-place E2E-V systems for binding government elections:

- The use of Scantegrity II (http://en.wikipedia.org/wiki/Scantegrity) in Takoma Park MD is still the only example of true E2E-V to have been used in a binding government election in the U.S. It combined E2E-V with familiar opscan-like paper ballots.
- A version of the Prêt à Voter E2E-V voting system was used for part of the State government election in the Australian state of Victoria, in November 2014.
- The StarVote project, which is expected to be used for binding government elections in Travis County and Dallas County, TX, beginning around 2018, also combines a paper evidence trail with an E2E-V voting system. Methods for facilitating risk-limiting audits of the paper trail are incorporated as well.

Some Internet E2E-V systems:

- Remotegrity was the remote voting solution for Takoma Park MD. Voters received a code sheet by postal mail, which then allowed them to cast their votes over an E2E-V Internet voting system linked to Scantegrity II.
- The Helios E2E-V Internet voting system (vote.heliosvoting.org) has been used in numerous non-government elections.

Non-E2E-V Internet voting systems may permit some verification of some steps. For example:

- Civitas (http://www.cs.cornell.edu/projects/civitas/) is a coercion-resistant Internet voting system that provides openly verifiable evidence that all the votes are correctly included and accurately tallied. However, it does not currently allow voters to verify that their votes are cast as they intended.
- The now-discontinued Norwegian Internet voting system incorporated an elegant code-based system for voters to check that their votes were cast as they intended. It also provided some evidence to some restricted observers that the outcome was tallied as recorded.

As far as we know, no end-to-end verifiable Internet voting scheme has been used for a large-scale binding government election. There is almost no experience with E2E-V in governmental elections, even in supervised settings.

## Conclusion and discussion of remaining issues

End-to-end Verifiable election technologies can be used in paper-based systems and in fully electronic systems, in remote voting systems and in-person poll-site systems, with simple majority counting methods and with many complex preferential schemes. *Any* electoral system can benefit greatly from the inclusion of E2E-V technologies. That said, the most vulnerable systems have the most to gain. There are good safeguards available for an election that is conducted entirely in-person and on hand-counted paper using simple rules. However, without E2E-V technologies, it is much more difficult to mitigate the inherent risks associated with remote and/or electronic systems. In particular, the risks of Internet voting systems are very substantial and they should not be contemplated without the mitigating benefits of E2E-verifiability.

However, while E2E-V appears to be a necessary condition for Internet voting, E2E-V Internet voting does not currently provide the same standards of evidence as E2E-V elections in supervised settings. We do not yet have much experience with E2E-V governmental elections in supervised settings, where procedures and the use of paper and statistical audits can help to address mishaps. The remote setting is far more challenging—especially without the use of a second channel (such as the postal service) —and also much less predictable. The only responsible path forward for Internet voting is to implement polling-place E2E-V voting and gradually experiment with more ambitious deployments without weakening the standard of evidence.

Remaining unsolved challenges include:

- ensuring that enough voters perform the cast-as-intended verification process properly,
- achieving dispute resolution without supervision of the voting process,
- achieving usable cast-as-intended verification without paper or some other similar material that allows the device to commit to an encrypted value.

There are partial solutions in each case. Remotegrity has some dispute resolution, but relies on a paper code sheet sent in conventional mail. Helios allows for cast-as-intended verification without paper, but only if the voter is very careful to record the encrypted votes before they are challenged. Some forms of encouragement for cast-as-intended verification could be performed remotely.

Overall, we are *not* asserting that Internet voting is viable if it is E2E-V, nor that the resulting scheme would achieve the same degree of privacy or verifiability as in-person (attendance) voting.